# Using a high-stability quartz-crystal microbalance to measure and model the chemical kinetics for gases in and on metals: oxygen in gold.

Alan J. Slavin

Department of Physics and Astronomy, Trent University, Peterborough, ON K9L 0G2, Canada

Electronic mail: aslavin@trentu.ca

This paper describes the use of a high-stability quartz-crystal microbalance (QCM) to measure the mass of a gas absorbed on and in the metal electrode on the quartz oscillator, when the gas pressure is low and the gas can be considered as rigidly attached to the metal, so viscosity effects are negligible. This provides an absolute measure of the total mass of gas uptake as a function of time, which can be used to model the kinetic processes involved. The technique can measure diffusion parameters of gases in metals close to room temperature at gas pressures much below one atmosphere, as relevant to surface processes such as atomic layer deposition and model studies of heterogeneous catalysis, whereas traditional diffusion measurements require temperatures over 400 $^o$C at gas pressures of at least a few Torr. A strong aspect of the method is the ability to combine the "bulk" measurement of absorbed mass by a QCM with a surface-sensitive technique such as Auger electron spectroscopy in the same vacuum chamber.

  The method is illustrated using atomic oxygen, formed under $O_2$ gas at 6 x 10$^{-5}$ Torr in the presence of a hot tungsten filament, interacting with the gold electrode on a QCM crystal held at 52 to 120 $^o$C. Some of the incident oxygen forms a surface oxide which eventually blocks more uptake, and the rest (about 80%) indiffuses. Surprisingly, the rate of oxygen uptake initially increases with the amount of oxygen previously absorbed; therefore, the measured oxygen uptake with time is reproducible only if pre-adsorption of oxygen conditions the sample. Temperatures above 130 °C are necessary for measurable thermal desorption, but all the oxygen can be removed by CO scavenging at all temperatures of these experiments. Simple kinetic models are developed for fitting the experimental QCM data to extract parameters including those controlling adsorption of oxygen, the CO scavenging probability, and limits on the diffusion jump frequency of dissolved oxygen. The reproducibility of the data and the good model fits to it provide proof-of-principle for the technique.

## I. INTRODUCTION

Diffusion of gases in metals is traditionally studied by measuring the quantity of gas passing through a metal diaphragm held at temperatures over 400 °C at gas pressures approaching 1 atmosphere or more, or measuring the desorption from a bulk sample saturated under similar conditions. (Ref. 1, pp. 3-8) For example, Outlaw et al.[2] measured the diffusion of oxygen through a silver diaphragm at 10 to 200 Torr between 400 and 800 °C. However, many processes potentially affected by gas diffusion, such as atomic layer deposition[3] and model studies of heterogeneous catalysis,[4] often occur close to room temperature at gas pressures much below one atmosphere. The experiments described here were carried out with the adsorbent (the gold electrode on a quartz-crystal microbalance) held between 52 and 120 °C, at an $O_2$ pressure of 6 x $10^{-5}$ Torr in the presence of a hot tungsten filament to create atomic oxygen.[5] This 6-MHz QCM gives an absolute measure of the mass of the adsorbate (oxygen), with a measurement stability of about 0.1 Hz over the 4 hours of an oxygen uptake run, equivalent to 3 % of a monolayer of oxygen. (Ref. 6 plus improvements described below). This is achieved primarily by controlling the sample temperature to ±0.02 K.

In comparison, commercial QCMs designed for thin-film deposition monitors usually use a water-cooled holder for the crystal, with the water temperature controlled to about +/- 1° C.[7] However, at 30 °C from the crystal's frequency-time (*f-t*) turning point, the variation in *f*(*T*) is of order 6 Hz/K, which means that the current QCM is some 50 times more stable than commercial devices as normally used. This opens the possibility of measuring the uptake of relatively light gases over extended periods of time, including the diffusion of gas into the electrode.

A QCM has also been used[8] to measure adsorption of $CO_2$, $CH_4$ and $N_2$ in small-pore metal-organic framework materials, but the mass of gas absorbed was at least 50 times that of the current experiments. Those experiments used a second gold-plated QCM as a reference sample to minimize the effect of temperature and pressure changes, but such an approach was not possible in the current work because the adsorption was onto gold. A QCM has also been used to measure absorption isotherms of water and various organic compounds from the gas phase, at gas partial pressures of 0.1 to 20 torr.[9] In this regime, the pressure of the gas has a significant effect on the QCM frequency, so an inert gas was added to keep the total pressure constant at 470 to 800 Torr. The sample chamber was contained within an air thermostat, so the inert gas also helped maintain the sample temperature constant. However, the experiment had no provision to measure the sample cleanliness during the experiments. An air thermostat was not needed in the current study because the temperature was controlled by heating the crystal holder. A QCM has previously been used to measure mercury vapour uptake in gold,[10] but this involved total frequency changes of order 1 kHz compared to less than 10 Hz in the current study.

The surface cleanliness of the gold in our experiments has been monitored by Auger electron spectroscopy (AES), allowing bulk and surface information to be obtained in the same vacuum chamber. This study is a proof-of-principle demonstration of the usefulness of the high-stability QCM as a tool in studying the adsorption and dissolution of gases in metals, and of the interaction of other gases (here CO) with this surface, both as a function of time, which enables kinetic modelling of the processes.

Since Haruta and co-workers[11] showed that nanoparticles of gold could act as effective catalysts for oxygen reactions, much effort has gone into elucidating the mechanisms involved. For example, Deng *et al.*[12] have studied the selective oxidation of styrene on O-covered Au(111) and thus have shown that extended surfaces of gold can also exhibit catalytic activity. Gold near room temperature is chemically inert to most common gases, except for the halogens. No adsorption of molecular $O_2$ on single-crystal gold surfaces has been observed for $O_2$ pressures to 1400 Torr and sample temperatures between 300 and 500 K.[13] However, Au can be oxidized below about 500 K, where oxygen desorbs, by several mechanisms (see references in Baker et al.[14]). One of these is the method used in this study, exposing $O_2$ to a hot filament (at least rhenium[15], platinum[16], and tungsten[5]) placed near the sample, which dissociates the molecule to atomic oxygen

The density of atoms in one Au(111) layer, $N_{Au} = 1.39 \times 10^{19}$ atoms m$^{-2}$, will be referred to as 1 MLE (monolayer equivalent). The maximum quantity of oxygen adsorbed on gold at room temperature has been measured by others to be in the range of 0.8 MLE to 2 MLE, with a variety of Au single-crystal surfaces and measurement techniques including AES and x-ray photoelectron spectroscopy (XPS), and temperature-programmed desorption (TPD) as calibrated by AES or XPS.[13,14,17,18,19]

Several of these authors used TPD to study the desorption kinetics of $O_2$ from Au as a function of sample temperature. For a disordered surface, Canning et al.[16] found a single desorption peak near 660 K that moved to higher temperature for low coverages (<0.07 MLE), but was almost independent of coverage over 1.5 MLE, suggesting first-order desorption kinetics for surface oxygen. The latter was surprising as one expects second-order kinetics for associative desorption, which gives a TPD peak whose temperature decreases with increasing initial coverage.[20] A TPD peak independent of coverage but near 550 K, was also seen for Au(111) by others.[12,17,18] In contrast, on Au(110) Sault et al.[13] found a TPD peak near 590 K with desorption kinetics that progressed from second order through first order and then zero order, as coverage increased. Min et al.[21] used primarily scanning tunnelling microscope (STM) and XPS to study the growth of oxygen overlayers on a Au(111) crystal and the rate of oxidation of CO to $CO_2$ on these surfaces. For their oxygen deposition at 400 K, the most relevant temperature for the current paper, large 2D islands consisting of a single oxide formed. Their study suggests that the oxidation of the CO occurs at the periphery of holes in this oxide. Baker et al.[14] have used STM images, TPD measurements, high resolution electron energy loss (HREEL) data and theoretical calculations. Their study revealed two different types of O present on Au(111) at high coverage: surface oxide and subsurface oxide. Shi and Stampfl[22] carried out a density functional calculation of oxygen on the Au(111) surface, and found that the most energetically favourable structure was a thin oxide-like one with an oxygen density of 0.3125 MLE. Gottfried et al.[23] and Baber et al.[24] studied oxygen-sputtered gold which also yielded some subsurface implanted oxygen. However, they saw up to four TPD peaks rather than the one dominant peak seen with the surface adsorption of oxygen. This indicates oxygen species unique to the sputtering process, so their results will not be considered further.

## II. EXPERIMENTAL

### A. Experimental methods

The experiments were carried out using a high-stability QCM[6] in a surface-science vacuum chamber equipped with a 750-eV argon ion gun for sample cleaning, AES with a LEED-type energy analyzer to identify surface species, and an MKS627B Baratron capacitance manometer model U5T for absolute measurements of gas pressure, with a full-scale range of 0.05 Torr and resolution of about $1 \times 10^{-7}$ Torr. The Baratron's zero-pressure reading drifted slowly during an experiment, which restricted the accuracy of the pressure measurements to about $1 \times 10^{-6}$ Torr.

The QCM crystal was positioned centrally in the chamber on a vertical xyz-rotation manipulator. The front of the crystal could either face the AES analyzer while spectra were being recorded, or be rotated through 180º to face the ion gun and the W filament during ion cleaning. The W filament was about 15 cm from the QCM crystal. A metal shield was rotated between the crystal and the AES screen during ion bombardment to protect the AES grids, and between the crystal and the W filament, during oxygen uptake or CO scavenging, to avoid the deposition of tungsten oxide and the heating of the crystal by the filament. The gas input ports were not in line-of-site of the crystal. See the apparatus schematic in **Supplementary Material**, section A, at (Add URL). Total pressures below $1 \times 10^{-6}$ Torr were measured with a Bayard-Alpert gauge, and partial pressures with a quadrupole mass spectrometer. The elastomer seal on the Baratron limited the base pressure of the chamber to $2 \times 10^{-8}$ Torr but gold is sufficiently inert that AES did not detect any significant contamination on the sample after several days held at this pressure. The QCM crystals, except where noted otherwise, were 6-MHz, AT-cut, plano-convex crystals, purchased from Colnatec (USA), but without the usual Cr or Ti bonding layer next to the quartz because Cr and Ti diffuse through Au[25,26] above 200 ºC and oxygen could bond to the dissolved Cr or Ti atoms. For good frequency control, it is important to obtain crystals with very few activity dips caused by interference with unwanted resonance modes, [27] and the Colnatec crystals were excellent in this respect. All the QCM crystals had a matte surface with a 7-micron finish, except as noted later, and a frequency-temperature (*f-T*) turning point near 85 °C. The average thickness of the as-received Au film was 170 nm according to the supplier, corresponding to 720 Au(111) monolayers assuming lattice close-packing on average. The QCM Au surfaces were of unknown crystalline structure, but other studies have shown that the crystallites composing them have primarily (111) surfaces,[28] so the computer modelling in this work is based on this assumption. Assuming other orientations should not alter the models greatly, but the modelling does ignore the possibility of diffusion at grain boundaries, which can be significant in Ag, for example.[2] All temperatures given below refer to that of the QCM crystal holder unless stated otherwise. The temperatures of the holder and the crystal were essentially the same, as shown by pressing a thermocouple against them.

As with most samples introduced into an ultrahigh vacuum system, a new QCM crystal was shown by AES to have a layer of carbon on the surface. This carbon prevented the uptake of any oxygen, and calls into question at least some of the diffusion results obtained by traditional methods that do not have the capability of measuring surface contamination or cleaning the surface in the same apparatus. The carbon also prevented oxygen adsorption on the rear electrode of the QCM crystal, which simplified the analysis. The front surface of a QCM crystal was cleaned initially, after installation from air, by argon-ion sputtering to remove typically 6 Au(111) monolayers followed by heating to 262 °C for 30 min to desorb any oxygen. The sample then showed only Au AES peaks from 20 to 600 eV with the Au 67-eV peak having its normal strength for a clean crystal. Once the sample was cleaned, it remained clean for several weeks at the system base pressure. All removal of Au by ion bombardment was measured by the QCM to keep track of the thickness of the Au film.

The QCM temperature controller measured a sample temperature stable to ±0.02 K, using a platinum resistance thermometer (RTD) attached to the QCM crystal holder.[6] Details on the temperature and pressure control, and the W filament, are given in the **Supplementary Material,** section A, at (**insert URL**). For perspective, 0.1 Hz is the inherent level of uncertainty of the frequency change over the entire course of an oxygen-uptake experiment, which typically gave a total frequency change of 8 Hz over 4 h.

The apparatus has been used for oxygen uptake runs from 52 to 120 °C, above which the *f-T* curve of the QCM crystal becomes so steep that even slight changes in sample temperature made the data unusable. A wider range of temperatures would be possible by using quartz crystals cut with different *f-T* turning points where the frequency is most stable. The position of the QCM crystal cannot be changed during measurement of oxygen absorption/desorption, to avoid frequency changes of several Hz caused by slight changes in the inductance or capacitance of the wiring of the quartz crystal. Once the crystal reached its desired temperature and just before exposure to oxygen, the vacuum chamber was tapped lightly a few times to release any pent-up stresses in the wiring. However, occasionally an abrupt jump in frequency up to 0.05 Hz would still occur during measurements. Such a jump was easily detected by eye in the graphed output, and was corrected for by subtracting the change from all subsequent data.

Introduction of $O_2$ or CO gas near 6 x $10^{-5}$ Torr caused an abrupt decrease in frequency, from about 0.01 Hz at 42 °C to 0.3 Hz at 120 °C, with corresponding increases when the gas was removed. It has been shown that these shifts are due largely to stress caused by differential thermal contraction between the QCM crystal and its holder.[29] The shifts cannot be caused primarily by the intrinsic temperature dependence of the quartz crystal, because the shifts are in the same direction above and below the *f-T* turning point. Although minor compared to the frequency changes from oxygen adsorption, these abrupt shifts have been subtracted in the data used in this paper. Useful TPD measurements could not be made: oxygen desorbing from the QCM holder totally overwhelmed that from the sample.

## B. Experimental results and discussion

### *1. AES results*

The top curve in Fig. 1 shows an AES spectrum of the oxygen signal on the gold film at near-saturation by oxygen, after 6000 s of $O_2$ exposure at 6 x $10^{-5}$ Torr with the sample at 85.2 °C, giving a frequency change of 8 Hz. The lower curve is the oxygen spectrum, taken at the same AES settings, from aluminium foil oxidized as-received, with enough ion bombardment to remove surface carbon. The lineshape of O on Au is similar to that for graphitic carbon and for SiC, where the bond is primarily covalent, rather than like the multi-lobed Auger structure for C in in $Ni_3C$, for example, which is primarily ionic.[30] Similarly, the two lower-energy peaks typical of $O^{2-}$ ions (closed $L$ shells)[31] that are seen in the $Al_2O_3$ spectrum in Fig. 1 are missing from O on Au, all suggesting that the oxygen-gold species on the surface is primarily covalent. The AES peak for O on Au is at 517 eV, compared to 507 eV for $Al_2O_3$. There was no measurable change in the position of the 67-eV Au AES peak up to near-saturation oxidation, although its amplitude decreased to about 0.7 of its clean value because of the overlying oxygen. The similarity in the AES strengths for the two curves in Fig.1 suggests comparable concentrations of oxygen near the surface, but an absolute comparison cannot be made from the AES data because of the different lineshapes, and different electron backscattering strengths for Au and Al. There were no measurable Au peaks near 517 eV to provide an amplitude comparison with O.

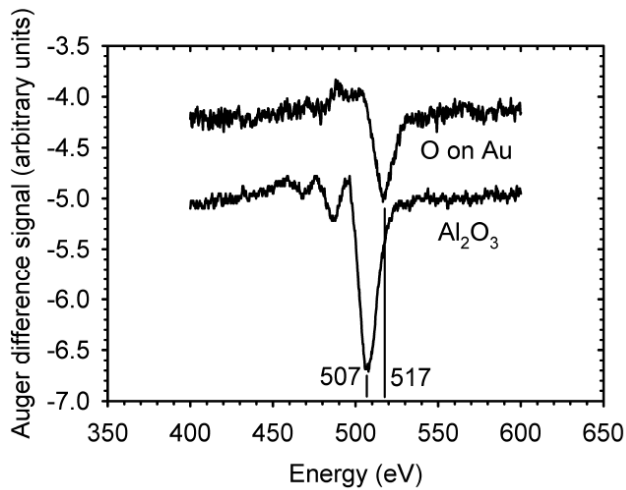

Fig. 1. Oxygen AES spectra. Upper curve: O on Au with the background from clean Au subtracted; lower curve: $Al_2O_3$. The peak minima are at 517 eV for O on Au, and 507 eV for $Al_2O_3$.

## 2. Uptake of oxygen measured by the QCM

The initial rate of oxygen uptake was measured to be proportional to the gas pressure, $P$, from $1 \times 10^{-5}$ to $10 \times 10^{-5}$ Torr. This is as expected for atomic adsorption. Adsorption of diatomic oxygen followed by dissociation would give a $P^{1/2}$ dependence.[32]

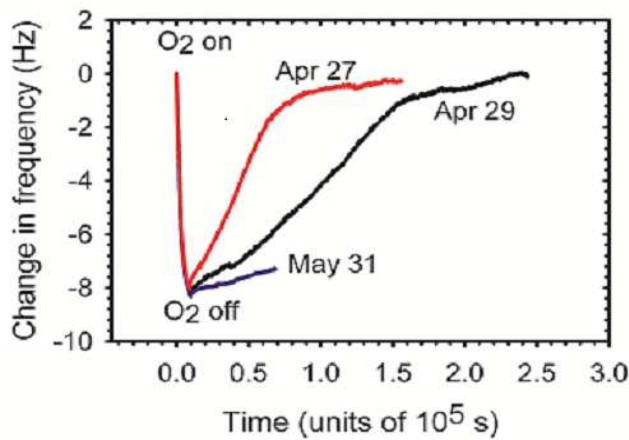

Fig. 2. (Color online) Three oxygen uptake curves for a sample held at 85.2 °C. The Apr 27 and Apr 29 data were taken immediately following ion cleaning. The May 31 data were taken 1 day after ion cleaning.

Fig. 2 shows the change in QCM frequency during exposure at 85.2 °C to $O_2$ at $6 \times 10^{-5}$ Torr for three different runs. The Apr 27 and Apr 29 data were taken 2 days apart, each immediately following an anneal of the QCM crystal to 260 °C for 15 min to remove residual oxygen, and then ion-cleaning. The May 31 data were taken 1 day after ion cleaning, with the sample held at 85.2 °C throughout. The oxygen-uptake parts of the curves (decreasing frequency) largely overlap on this time scale. For the first two runs, after the $O_2$ was turned off the oxygen immediately began to desorb rapidly, and this continued over the next two days until essentially all the oxygen was gone from the sample. However, the desorption rates varied by roughly a factor of 2 between the two runs. This rapid and variable desorption rate was typical for all runs for which the Au film was ion-cleaned without a subsequent anneal prior to adsorbing oxygen. In contrast, if the sample was annealed to 260 °C for 15 min after ion cleaning, or if the sample was left for at least a day in vacuum at over 85 °C before exposing to oxygen, there was much slower desorption, as seen in the May 31 data of Fig. 2, where the rise in frequency can be explained totally as scavenging of oxygen by background CO in the chamber which reacts with O atoms on the Au to form $CO_2$. The data of Fig. 2 suggest that the ion bombardment produced short-lived oxygen traps that could be annealed out, although scavenging by background CO would also have been occurring during this time. Therefore, all other data reported below were taken after the sample was ion-cleaned and then annealed to 260 °C for 30 min. As discussed later, the sample was eventually cleaned of oxygen entirely by CO scavenging; this did not require any ion-cleaning or annealing after the initial cleaning, and gave reproducible uptake curves. However, no CO was introduced intentionally during the time of the runs in Fig. 2, so any CO scavenging there is from background gas only.

The lower curve in Fig. 3, at 85.3 °C is the May 31 oxygen-uptake from Fig. 2. It is typical of those taken for a single exposure to $O_2$ at 6 x $10^{-5}$ Torr, after cleaning and annealing (or after CO scavenging). Points at which the $O_2$ or CO gas were turned on or off are labelled. The lower curve is initially slightly concave-down after the $O_2$ is admitted, showing an increase in the rate of oxygen uptake to about 5 ks, and then becomes concave-up as absorption slows. The upper curve, at 52.2 °C, shows initial oxygen uptake on a clean sample to a frequency decrease of about 2 Hz, followed by CO scavenging at 2.5 x $10^{-5}$ Torr for 1 ks (2.5 x $10^4$ langmuirs (L)) to remove all the oxygen, followed by a second oxygen uptake to near saturation, followed by a final CO scavenge at the same CO pressure. Both curves include the correction for room-temperature changes. The upper curve shows that, after CO scavenging, the frequency returned to essentially its value before oxygen was admitted. That is, there was no significant hysteresis in the crystal frequency during an experiment.

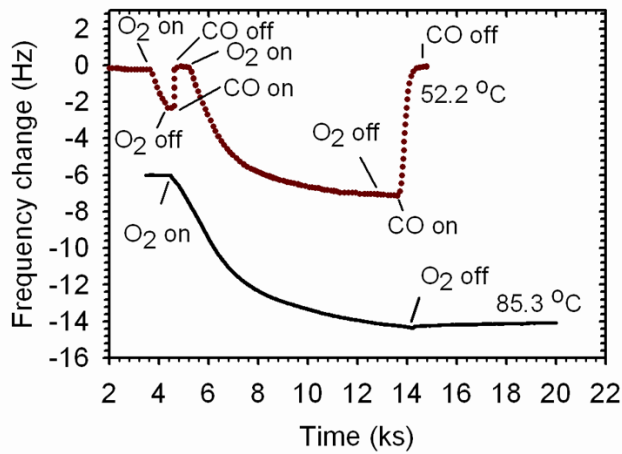

Fig. 3. (Color online) Change in frequency during oxygen uptake at 6 x $10^{-5}$ Torr. The top curve includes a conditioning absorption. The lower curve is without any conditioning absorption.

The initial adsorption rate for the second $O_2$ exposure in the upper trace in Fig. 3 was about 20% faster than for the first one, and the concave-down portion was absent for the second $O_2$ exposure. This general pattern was always observed with two $O_2$ exposures separated by a full CO scavenge. The first $O_2$ exposure somehow conditions the sample to allow the oxygen to adsorb on, or diffuse into, the sample more rapidly the second time. This explains the change from concave-up to concave-down in the lower curve without a pre-exposure: as oxygen is adsorbed, the sample is conditioned to adsorb more quickly, until eventually saturation is approached and adsorption slows. A similar increase in diffusivity has been observed by Shelby (Ref.1, p.21) for the out-diffusion of helium through vitreous silica after removal of the gas at high pressure. He described this behaviour as "strong evidence that the diffusivity is enhanced by the high concentration of gas dissolved in the specimen."…"This expansion will slightly enlarge the doorways and perhaps cause a temporary enhancement of the diffusivity". A similar mechanism may explain the results of Fig. 3, if the initial oxygen uptake causes a metastable rearrangement of the Au atoms that facilitates in-diffusion. Whatever the cause, the higher rate of oxygen uptake after an initial absorption/scavenging procedure was seen in every O uptake run measured with the QCM. The double oxygen adsorption process followed in the upper curve will be called a "standard" uptake run, and was essential for reproducibility of the oxygen-uptake curves; it was used for all data discussed below. The CO exposure used here for the final scavenge was well in excess of the 8000 L used by Gottfried et al.[23] to remove all TPD peaks below 600 K for the Au(110) surface.

The delay of about 1000 s between the first and second $O_2$ exposures in Fig. 3 was clearly small enough for the conditioning to remain in effect for the second exposure. Moreover, when a second standard uptake run was carried out beginning about 30 min after the first one, the initial uptake rate was the same as that for the first standard run, showing that the sample conditioning persisted for at least 30 min. However, if the standard run was repeated the next day, the uptake rate after the conditioning exposure was again 20 to 30% faster than the first one, showing that the conditioning of the sample persisted less than 1 day. Outlaw et al.[2] also required several initial runs for their study of the diffusion of oxygen through a Ag diaphragm before results became reproducible. They presumed that this was due to contaminant removal during the initial runs, but our work suggests that preconditioning of the Ag may also have contributed.

Figure 4 shows oxygen uptake curves, $m(t)$ in units of $10^{19}$ atoms m$^{-2}$, obtained at 85.2 °C from the frequency-time curves as corrected for room temperature changes, by multiplying the frequency change by - 4.63 x 10$^{18}$ atoms m$^{-2}$ Hz$^{-1}$. This multiplier was obtained as the crystal sensitivity in kg m$^{-2}$ Hz$^{-1}$ from the Sauerbrey equation,[33] divided by the mass of one oxygen atom. Although the QCM crystals are most sensitive near their centres, it has been shown that for uniform surface coverage, as in the current experiments, the Sauerbrey equation gives the correct mass value within 1%.[34] To show the degree of reproducibility, the four upper lines in Fig. 4 are all for standard runs, taken over 1 month. These curves coincide within the uncertainty of the measurements, although one curve (Jul19) taken a month after the other three is slightly lower. The lowest curve in Fig. 4 shows an O uptake curve without a preceding oxygen uptake, and is significantly different from the others. The individual data points in Fig. 4 are oxygen AES amplitudes and will be discussed later.

.

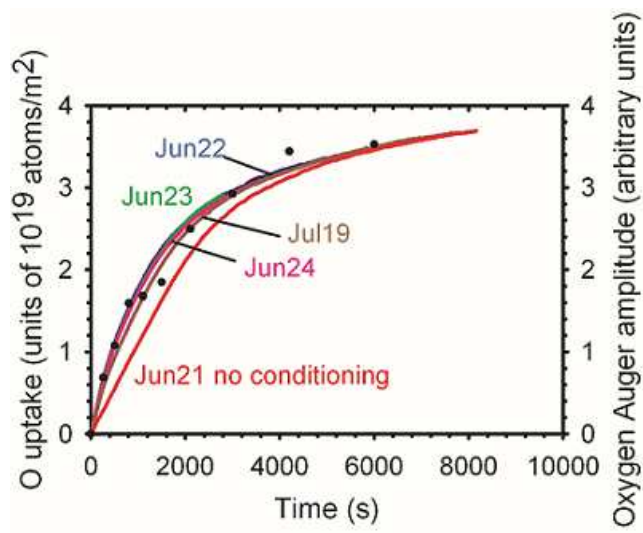

Fig. 4. (Color online) Lines: Oxygen uptake at 85.2 °C, measured by the QCM. Dots: Auger amplitude during oxygen uptake. The Jun23 line almost coincides with the Jun22 one. The Jun21 line was for no pre-conditioning.

The total uptake of oxygen in the gold film was typically $3.7 \times 10^{19}$ atoms m$^{-2}$ near maximum coverage (referred to here as "near-saturation"), or 2.7 MLE. For comparison, the oxygen required to form a single layer of PbO on a Pb film evaporated onto a Au QCM electrode produced a QCM frequency shift of 1.75 Hz, or 0.58 MLE.[35] (The PbO oxygen AES lineshape was triple-lobed like the $Al_2O_3$ one in Fig. 1 so a direct comparison of quantities cannot be made from the AES results.) Clearly, all the 2.7 MLE of absorbed O on Au cannot be accommodated on the surface. It will be argued later that much of the oxygen is dissolved in the Au below the surface, mostly beneath the AES probing depth.

## 3. Uptake of oxygen measured by AES

The individual data points in Fig. 4 are oxygen AES amplitudes from a standard oxygen adsorption run done at 85.2 °C under $6 \times 10^{-5}$ Torr of $O_2$, with the oxygen exposure interrupted at intervals and the AES signal recorded to determine the quantity of surface oxygen, for comparison with the QCM uptake curves. The background Auger signal for a clean Au sample was subtracted from each of the AES curves, and then the height of the downward lobe of the O peak was measured relative to the curve between 530 and 550 eV where the difference spectrum was flat. The AES data (right axis in Fig. 4) has been normalized arbitrarily to the oxygen-uptake axis at about 2000 s. The shapes of the two curves are very similar, suggesting that the total oxygen uptake is being limited by the formation of surface oxygen. Henceforth, this surface oxygen will be referred to as "oxide", based on the experimental and theoretical work of others.[14,21,22]

## 4. Evidence that the large QCM frequency change is due to oxygen absorption and not caused by spurious effects

Our QCM result, that about 2.7 MLE of oxygen has been absorbed by the Au film, is significantly higher than the maximum 2 MLE estimate by other researchers from AES and XPS. We here dispense with other possible causes of a frequency shift that might explain an anomalously large QCM frequency shift: (a) the adsorption of a gaseous impurity during $O_2$ exposure, (b) stress on the QCM crystal from the surface oxygen, and (c) a surface area considerably larger than the perpendicular area of the QCM electrode, because of the 7-micron matte finish.

(a) The only impurity, apart from oxygen, seen in AES over the energy range 20 to 600 eV was a small signal from Cl at 181 eV, typically about the same strength as the Au 150-eV peak. A comparison of these two peaks in the Handbook of Auger Electron Spectroscopy[36] shows that the AES signal from Cl is about 60 times stronger per atom than for the Au. This implies a Cl coverage of roughly 0.02 MLE on the gold surface, corresponding to a frequency change of only about 0.1 Hz. This cannot explain the difference between the oxygen seen at the surface in AES or XPS, and the total O absorbed as measured by the QCM. Moreover, the Au 67-eV AES peak was not measurably reduced by the presence of this small Cl peak, supporting the conclusion of negligible surface contamination.

(b) Stress in the electrode of a QCM crystal can cause significant frequency shifts.[37] However, gold is a soft metal with low elastic moduli, so it is unlikely that the stress from oxygen on the surface could have propagated through the Au to affect the quartz crystal frequency. This conclusion is supported by the saturation oxidation of a Pb film on the Au electrode of a QCM crystal which caused a frequency shift of only 1.75 Hz that was totally explicable by the formation of stoichiometric PbO on the surface[35] with no evidence of a stress effect. Similarly, oxidation of Sb films on a Au QCM electrode produced frequency shifts that were totally explicable by stoichiometric oxide compositions.[38] Finally, an absorption run was also carried out using an RC crystal from Phillip Technologies USA, which is claimed to have a very low sensitivity to film stress. There was no substantial difference in the total frequency shift between that run and runs using AT-cut crystals, although the larger dependence of frequency on temperature of the RC crystals (about 14 Hz/°C) made them less useful than the AT-cut ones for our purposes. Therefore, any frequency change from surface stresses is expected to be negligible in the current experiment on the adsorption of oxygen.

(c) The 7-micron finish on the crystals used in these experiments will increase the absorption area and so could increase the mass of oxygen adsorbed over that measured on polished surfaces. However, Knutson[39] deposited electrolytically the same amount of metal on polished and matte QCM crystals, and obtained the same frequency-to-mass factor for both within 1.5%, showing that the actual areas are very similar. Moreover, an oxygen absorption run was carried by us out with a polished RC crystal (Phillip Technology, USA) with a 0.01-μm finish. It gave about the same frequency shift from O absorption as for the matte surface used in the rest of the experiments.

Finally, as discussed later, CO scavenging was used to remove only about 2/3 of the total absorbed O, as measured with the QCM. At this point, the O AES signal was reduced to about 20% of its original value, close to the noise level, although the QCM frequency showed that about 1 MLE of O remained in the sample. That amount of oxygen would have produced a large AES signal if on the surface. This is strong proof that the remaining O was below the surface of the Au. An STM and XPS study has shown that at 400 K, oxidation of the Au(111) surface forms large 2D islands consisting of a single oxide.[21] It is henceforth assumed that the remaining oxygen was dissolved in the Au rather than forming a subsurface oxide. This assumption will be considered again later.

# III. COMPUTER MODELLING OF THE QCM DATA TO EXTRACT KINETIC PARAMETERS

## A. Introduction

If the dissolved oxygen were in equilibrium with the surface oxide observed in AES then, if the surface oxygen is largely removed by CO scavenging but not all the dissolved oxygen, annealing the sample at an elevated temperature in vacuum should allow dissolved oxygen to segregate back to the surface to form oxide until equilibrium is re-established. However, after about 2/3 of the oxygen was removed to reduce the O AES signal to about 20% of its original value, heating the crystal over an hour in about 10-degree steps from 104 °C to 193 °C, with AES measurements at each step, did not show any increase in surface oxygen using AES. A further anneal at 115 °C for 30 minutes also showed no increase in surface oxygen. More generally, no increase in the O AES signal was ever observed after annealing the sample in vacuum at any temperature up to 190 °C, even with annealing times of a few days.

The conclusion is that the dissolved oxygen is not in equilibrium with the surface oxide, and these two components occur by different pathways. That is, when the oxygen atoms strike the surface, some of them form surface oxide and some immediately dissolve, and the dissolved oxygen does not subsequently surface-segregate to form oxide. This leads to the computer modelling in the following sections. All the data-fitting described here was carried out on the Sharcnet.ca Canadian computing network.[40]

## B. Removal of all the oxygen by CO scavenging: qualitative discussion

The sample was cleaned of oxygen after each of the standard-run oxygen-uptake experiments discussed below by exposing it to CO gas. The QCM showed that all the absorbed oxygen – both surface oxide and dissolved oxygen – could be removed by the CO as seen in the top curve of Fig. 3; the surface was then also clean as measured by AES, and this avoided the need to sputter-clean the crystal and anneal it after each oxygen uptake experiment. In principle, QCM measurements as a function of time during CO scavenging of the oxygen should allow modelling to provide a measure of the jump frequencies for O atoms between adjacent Au layers, as well as values for the scavenging probabilities for the oxide and dissolved oxygen. Fig. 5 shows two full CO scavenging curves at the same temperature, 52.2 °C, but quite different CO pressures: 25 µTorr and 121 µTorr, as labelled.

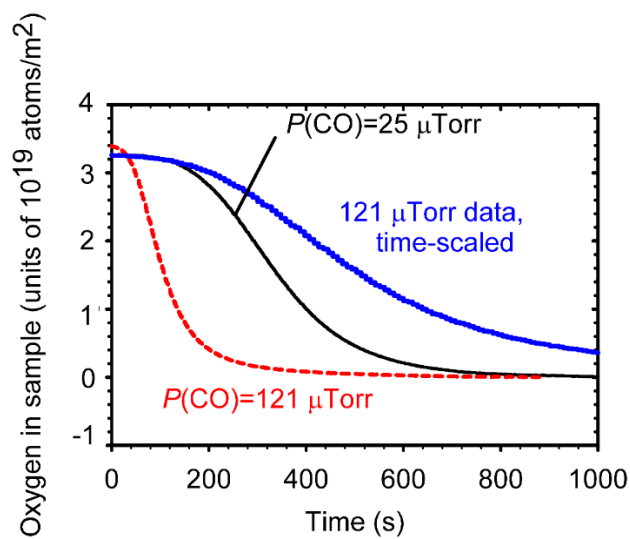

Fig. 5. (Color online) Complete CO scavenging of oxygen after uptake at 52.2 °C. The CO was admitted at 0 s. The right-hand (blue) curve is obtained from the 121-$\mu$Torr one by scaling the time axis by 121/25 and normalizing the starting amplitude to the 25-$\mu$Torr value.

The starting amount of absorbed oxygen in the two runs in Fig. 5 was very similar and near saturation, so the degree of initial oxide coverage should be similar. If the scavenging rate were limited by out-diffusion of O atoms rather than by the incident CO flux, then the time for total depletion should be similar for these two curves, which is not the case. Alternatively, if the scavenging rate just depended linearly on the incident CO flux density (assuming a very high rate of out-diffusion) then one would expect these two curves to be displaced by a time factor of 121/25 = 4.8. This time scaling of the dashed curve is shown as the (blue) curve on the right, whose amplitude has been normalized to that of the 25-µTorr one at $t$ = 0 s. The right-hand curve falls much more slowly than the 25-µTorr one after about the first 150 s, indicating that other factors are affecting the scavenging rates. Clearly, the oxide and dissolved oxygen must be considered together to explain these experimental results.

The results of Fig. 5 show that the scavenging rate initially started off very slowly with a parabolic-like decline: after the admission of CO gas, it could take over 30 s before there was a measurable (0.02 Hz) decline in the QCM frequency, which then accelerated rapidly. The most likely reason for this follows from the suggestion by Min *et al*.[21], that the scavenging of the oxide occurs at the periphery of holes in the oxide. As these holes expand, the scavenge rate would increase until the holes coalesce, after which the oxide periphery would decrease and so would the oxide scavenging rate. Regardless of the initial scavenging mechanism, the scavenging rate must slow eventually, because the probability that an incoming CO molecule will strike an oxygen atom decreases as both the oxide coverage and the amount of dissolved oxygen decrease, and because some of the dissolved oxygen must make its way in a random walk to the surface before it can be scavenged. The modelling of the entire CO scavenging curve as in Fig. 5 is discussed later.

Parker and Koel[18] have also measured the dependence of the O AES strength on CO scavenging time for a Au(111) surface, but their results only relate to the surface oxide, not the dissolved oxygen. They found that all the surface oxide was removed in about 1200 s at a CO pressure of $5 \times 10^{-8}$ Torr. It is unclear why it took almost as long to remove all the oxygen from our sample at the much higher CO pressure of 25 µTorr, even with both oxide and dissolved oxygen needing to be removed.

## C. Scavenging of oxygen by background CO

One must ask if background CO in the chamber affected the oxygen uptake measurements such as those in Fig. 4. A background CO partial pressure $P(CO)$ of about $1.5 \times 10^{-6}$ Torr was measured with the mass spectrometer when $O_2$ was present at $6 \times 10^{-5}$ Torr. Some CO is to be expected, because CO is produced in $O_2$ gas at hot W filaments.[41] However, the measured $P(CO)$ remained unchanged when the W filament was turned off, showing that the measured CO was being generated predominantly by the filament in the mass spectrometer head. Nevertheless, the background CO under vacuum still gave a $P(CO)$ of typically 1.5 x $10^{-8}$ Torr, so some CO was also clearly present during $O_2$ exposure. The O uptake curves, as in Fig. 4, became nearly linear when the sample was close to full oxide coverage, with a slope of typically 8 x $10^{14}$ atoms $m^{-2}s^{-1}$. It is likely that background CO continuously scavenged some surface oxide, preventing total coverage by the oxide during uptake and allowing more dissolved O to enter the Au. This would create the observed linear tail of the uptake curves.

Background CO can also explain the slow increase in QCM frequency of about 1 x$10^{-5}$ Hz $s^{-1}$, or about 5 x $10^{13}$ atoms $m^{-2}s^{-1}$, after an O uptake run when the sample was kept in vacuum overnight, as in the lowest curve of Fig. 2. This slow increase was not caused by thermal desorption: from Section IV, the rate of thermal desorption at 85 °C is too slow to explain the desorption in Fig. 2. If the probability that a CO molecule strikes an O atom in the oxide were given by the fractional O surface concentration $\theta$, then the results of Fig. 2 would provide a rough estimate for the scavenging probability $a_{ox}$ of an O atom from the oxide by an incident CO molecule. However, it will be shown later that, at high oxide coverage, the CO scavenging occurs only at the peripheries of holes in the oxide, so the assumption of proportionality to $\theta$ is incorrect.

One observation remains unexplained: why does scavenging of a nearly O-saturated surface in Fig. 2 by background CO cause an approximately linear increase in the QCM frequency, whereas the scavenging curve is more quadratic to start in Fig. 5? The CO pressure is 3 orders of magnitude greater for the results of Fig. 5, and this may cause a change in the scavenging kinetics.

## D. Variation of AES signals as the oxygen is scavenged

It was originally hoped that the effects of surface oxide and dissolved oxygen in the CO scavenging curves could be separated by exposing the sample to enough CO to remove only the oxide layer. To check this possibility, Fig. 6 shows the Au (67 eV) and O (517 eV) AES amplitudes as a function of the amount of oxygen (in Hz) scavenged after a standard absorption run at 85.1 °C. The O amplitudes were measured as discussed earlier, and the Au amplitudes were measured peak-to-peak with background subtraction not being possible because the Au peak was always present and large. The measurements were made starting with a near-saturation oxygen uptake of 8.0 Hz. After this, the sample was subjected to a series of CO exposures at a measured pressure to remove about 1 Hz of oxygen each time, followed by the AES measurement. The crystal frequency was then allowed to stabilize before the next CO exposure, so the frequency change from scavenging could be measured. This continued until the O AES signal was no longer measurable. By the time 5 or 6 Hz of O was scavenged, the O AES peak was reduced to about 20% of its initial height, or only about twice the noise level. The solid lines in Fig. 6 are linear fits through the data points.

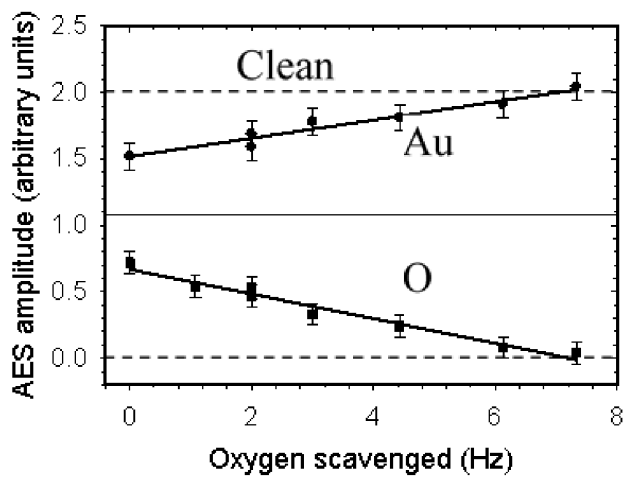

FIG. 6. AES amplitudes as a function of the amount of oxygen scavenged, at 85.1 °C.

From Fig. 6 and several other experiments, after about 5 Hz of oxygen was scavenged following near-saturation coverage there was still about 1.6 MLE of oxygen in the sample, even though the AES peak had been reduced to 15 to 20% of its original value. However, 1.6 MLE of oxygen at the surface would have produced a much stronger AES peak than observed here (based on PbO on Au)[42], under the same AES operating parameters. Given that the electron attenuation length is 6.6 Å for the O AES peak,[43] this implies that most of the remaining oxygen for O on Au must lie below the surface. Another conclusion from Fig. 6 is that some surface oxide was present until at least most of the oxygen was removed, which confirms that both the oxide and dissolved oxygen must be considered together to model the scavenging.

It seems likely that the dissolved O is located in a selvedge several layers thick, close to the Au surface. If the dissolved O were distributed uniformly throughout the roughly 700 Au(111) layers of the Au film, the near-surface concentration would be too low to observe in AES once the surface oxide was removed. Such a selvedge could be produced in two ways. (1) As discussed earlier, a pre-adsorption of oxygen appears to cause a rearrangement of the near-surface Au atoms to increase the rate of dissolution, but this would be expected to occur only close to the surface where the Au atoms have more room to rearrange. (2) It is possible that the bombardment during the ion cleaning also contributes to such a porous region. In this case, the thickness of the selvedge would be given by the stopping distance of the 750-eV Ar ions in Au, which can be calculated using the SRIM program.[44] This gives a value of 9.5 Å with a "longitudinal straggling" (standard deviation) of 14.5 Å. That is, the overall thickness of the bombardment selvedge is about 25 Å, or 10 Au(111) layers, which is the value assumed for the calculations below. The validity of this assumption will be considered later.

From the Einstein equation for a random walk, the root-mean-square displacement of an atom diffusing in a given direction in time $t$ is given by $d = \sqrt{2Dt}$, where $D$ is the diffusion constant. In cubic materials, $D = \lambda^2 f_i$,[45] so

$$f_i = (d/\lambda)^2 / (2t), \qquad (1)$$

where $f_i$ is the internal jump frequency between adjacent Au planes and $\lambda$ is the jump distance, here the (111) layer spacing. If the 121-µTorr curve in Fig. 5 were diffusion-limited, then it shows that an atom must be able to travel an average distance of at least $10\lambda$ (the assumed selvedge thickness) in about 200 s because the oxygen is almost all desorbed by this time. From Eq. (1), this gives a minimum jump frequency of roughly $f_i \sim 10^2/(2 \times 200) = 0.25$ Hz for this data.

## E. Modelling the CO scavenging curves

We model the CO scavenging curves for a near-saturation of oxygen by assuming that scavenging occurs only at the periphery of holes in the oxide, as suggested by Min *et al.*[21] (Other models were also tried, without a good fit to experimental data.) All the CO molecules landing in a hole, within an "access ring" of width $W$ inside a hole's perimeter, will reach the periphery, where $W$ is the average distance in a given direction that a CO molecule can travel during its residency time on the surface. Therefore, like the situation for the thermal desorption of atoms from the edges of 2-dimensional islands on a surface,[46,47] the holes will expand at a constant velocity $v$ perpendicular to the boundary.

The model variables were $W$, *aox, adis*, and $f_i$, where $W$ is the average distance in a given direction that a CO molecule can travel during its residency time on the gold surface, $f_i$ is an internal jump frequency between adjacent Au(111) planes, *aox* is the probability that an incident CO molecule reaching the oxide periphery desorbs as $CO_2$, and *adis* is the probability that a CO molecule striking a dissolved O atom desorbs as $CO_2$. However, the product of $W$ and *aox* always occurred together and so is denoted *W_aox* below, reducing the number of fitting parameters to three. The oxygen areal density in the oxide is taken as the theoretical value *oxdens* = $0.3125NAu$.[22] Even if the actual value is somewhat different from this, it should not affect the modelling results qualitatively. The full description of this modelling is found in the **Supplementary Material**, section B, at (**insert URL**). The computer codes are also in the **Supplementary Material,** section D. If $F_{CO}$ is the CO molecular flux density, then the value of $v$ is calculated as

$$v = F_{CO} * W\_aox / oxdens. \qquad (2)$$

The root-mean sum of squares of the differences between the model output and the QCM data was minimized using the Nelder-Mead Downhill Simplex procedure.[48] This rms sum was divided by the maximum value in the experimental uptake, to give a goodness-of-fit parameter *yrms* which will be about 0.01 for an average 1% deviation between experiment and model results.

A typical fit to the data is given in Fig. 7 for the Jul28 run at 52.2 °C and $P(CO) =$ 2.5 x $10^{-5}$ Torr, as well as the curves showing the calculated areal density of O atoms in the oxide and dissolved. Note that about 0.7 MLE of dissolved O remains when the oxide is all gone.

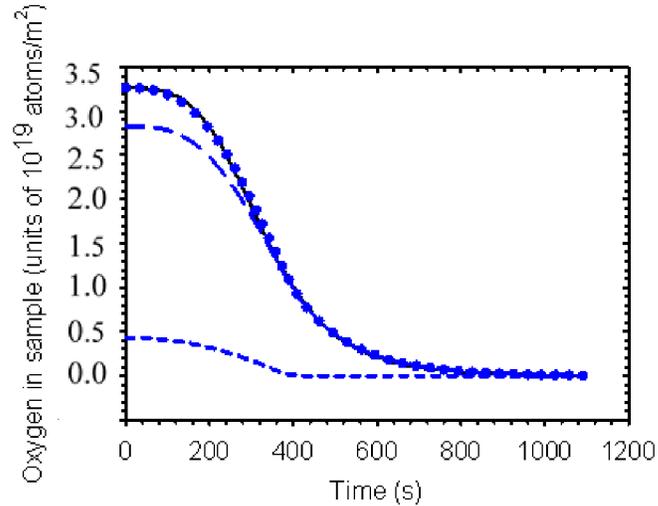

Fig. 7. (Color online) Data and fit for full CO scavenging of oxygen at 52.2 °C and $P(CO) = 2.5$ x $10^{-5}$ Torr. Solid line (black): experimental data. Dots: overall fit. Short dash: O in oxide. Long dash: total dissolved O.

Table 1 lists the values of the fitting parameters for all the runs subjected to complete CO scavenging, after near-saturation oxygen uptake. The velocity $v$ was calculated from Eq. (2). The necessity to set a constant CO pressure quickly for the scavenging means that the value could not be set reproducibly from run to run, so it varied from 2.5 x $10^{-5}$ to 13.6 x $10^{-5}$ torr in these runs. The data given in boldface in Table I, with $f_i$ listed as "infinite", was calculated assuming that the jump frequency was so large that the concentration of dissolved oxygen was the same in all layers of the selvedge, as discussed later.

Table I. Values of *yrms*, *W_aox*, *adis*, *fi* for CO-scavenging runs after maximum O uptake, at various sample temperatures and CO pressures. The numerical factors in the top row give the units for the corresponding column.

| T °C | Run | $P_{CO}$ $10^{-5}$ Torr | yrms $10^{-3}$ | W_aox $10^{-5}$ m | adis $10^{-3}$ | fi Hz | v $10^{-3}$ $R_c$/s | v/$P_{CO}$ $10^2$ $R_c$/Torr-s |
|---|---|---|---|---|---|---|---|---|
| 52.2 | Jul28 | 2.5 | 5.86 | 12.3 | 10.7 | 21.2 | 2.72 | 1.09 |
|  |  |  | **5.74** | **12.4** | **10.6** | **infinite** | **2.74** | **1.10** |
| 52.3 | Aug04 | 12.1 | 5.27 | 7.61 | 15.4 | 0.61 | 8.11 | 0.67 |
|  |  |  | **7.69** | **11.0** | **4.11** | **infinite** | **11.7** | **0.97** |
| 53.0 | Aug26 | 3.2 | 5.52 | 9.79 | 8.12 | 2.8 x $10^4$ | 2.76 | 0.86 |
|  |  |  | **5.43** | **9.86** | **8.05** | **infinite** | **2.78** | **0.87** |
| 101.1 | Aug05 | 11.0 | 6.83 | 11.5 | 8.20 | 7.0 | 11.2 | 1.02 |
|  |  |  | **6.73** | **12.1** | **7.22** | **infinite** | **11.7** | **1.06** |
| 101.1 | Aug08 | 10.9 | 8.22 | 11.8 | 8.21 | 2.6 x $10^4$ | 11.3 | 1.04 |
|  |  |  | **7.74** | **12.2** | **7.93** | **infinite** | **11.7** | **1.07** |
| 101.1 | Aug09 | 13.6 | 4.11 | 9.15 | 35.4 | 1.2 | 11.0 | 0.81 |
|  |  |  | **12.4** | **15.1** | **6.97** | **infinite** | **18.0** | **1.32** |
| 101.1 | Aug25 | 3.4 | 5.45 | 13.8 | 13.9 | 1.2 | 4.08 | 1.20 |
|  |  |  | **5.83** | **14.9** | **10.3** | **infinite** | **4.41** | **1.29** |
|  | **Average** |  | **7.6** | **12.9** | **7.7** | **infinite** |  | **1.14** |
|  | **St. dev.** |  | **2.2** | **1.9** | **1.9** | **infinite** |  | **0.17** |

All the fits that included *fi* (not boldface) agree with the experimental curves with *yrms values* better than 1%. There is no obvious temperature dependence of the fitting parameters over the range of 50 °C used: the averages for *W_aox* and *adis* at 52 °C and at 101 °C overlap within their standard deviations. The *W_aox* values agree within a factor of 2, and also the *adis* values except for the Aug09 run. The values for the velocity are as anticipated, to give coalescence near the inflection point in the scavenging curve; e.g., for the Jul28 run, $R_c/v$ = 368 s, consistent with the inflection point in Fig. 7. From Eq. (2), the velocity *v* is proportional to *P*(CO), so one expects *v*/*P*(CO) to be approximately constant over these runs if *W_aox* is independent of temperature. This is verified in the last column of Table 1: *v*/*P*(CO) varies from 0.67 x $10^2$ to 1.32 x $10^2$ with no apparent temperature dependence.

The *fi* values require more discussion. Most of the *fi* values are quite low (0.6 to 21 Hz). In contrast, the Aug26 and Aug08 runs have much higher jump frequencies: for the Aug28 run, $2.8 \times 10^4$ Hz at minimization of *yrms* and, for the Aug08 run, $2.6 \times 10^4$ Hz and still increasing when the optimization procedure was terminated. However, in both cases *W_aox* and *adis* converged to the values in the table, for *fi* values as low as 250 Hz. Comparing the fitting parameters for the Aug05 and Aug08 runs is instructive. These experiments were done at almost the same CO pressure and the QCM data were identical within about 1% over the entire scavenging time. The *W_aox* and *adis* parameters are also essentially identical, and the fitted results analogous to Fig. 7 overlapped within about 1%. The conclusion is that the model is very insensitive to the exact value of the jump frequency, which suggests that the diffusion is rapid on the time scale of these experiments, and it is the value of *P*(CO) that primarily determined the scavenging rate for the dissolved oxygen over the entire scavenging time, not the out-diffusion time. To check this, the computer model was modified for the Aug08 to assume an infinite jump frequency by making the areal density of dissolved oxygen uniform throughout the selvedge after each time step; this removed the diffusion component in the fitting code and reduced the number of fitting parameters to just two. This affected the fit for the Aug08 run very little and gave *W_aox* and *adis* values which were only slightly different from those for the 3-parameter model.

Based on this success, the 2-parameter model was applied to all the CO scavenging experimental data, with the results shown in boldface everywhere in Table 1 and with the *fi* column labelled "infinite". This 2-parameter model also fitted the experimental data within 1% except for the Aug09 run at 1.2%, and the values of *W_aox* and *adis* for the 3-parameter and 2-parameter fits generally agree well with each other with the main exceptions being the Aug04 and Aug09 runs which both had low *fi* values in the 3-parameter model. It is not obvious why *W_aox* and *adis* for the 3-parameter fit for the Aug25 run, which also had a low *fi* value, agreed reasonably well with values from the 2-parameter model. As expected, the calculated values for the areal density in a single layer of the Au was fairly uniform for the 3-parameter model with high *fi* values, but increased significantly with depth for runs with low *fi* values. The main conclusion from Table 1 is that the 2-parameter model is probably adequate for all the data, and the out-diffusion of the oxygen does not determine the overall scavenging time. The excellent fit of the modelled curve to the experimental data, using only 2 adjustable parameters, adds confidence for the model. The average and standard deviation from the mean for each of the fitted parameters, for the 2-parameter fit, are given in boldface in Table I. As for the 3-parameter model, there is no obvious temperature dependence.

We return now to the question of the thickness of the selvedge, assumed above to be 10 Au(111) layers. Fig. 7 is self-consistent with the assumption of a thin selvedge that contains the dissolved O, because an O AES signal was still visible when all the oxide had been scavenged according to Fig. 7, and this would not be possible if the O was distributed over many more than about 10 Au(111) layers of the film.

A check on the choice of selvedge thickness in the model is obtained by comparing the form of the AES results in Fig. 6 with a simulation using the fitted results in Fig. 7 for the amount of oxygen in the oxide or dissolved. The AES signal should be proportional to the areal density of oxygen atoms in the oxide, plus the density of the dissolved O as attenuated by the overlying gold. From the fit in Fig. 7, the dissolved oxygen was found to be distributed uniformly, within 1.5 %, over the $S = 10$ layers of the selvedge, so the O density per layer was $Totd/S$ where $Totd$ is the total dissolved oxygen. The transmission factor for oxygen through one overlayer of Au is $\exp(-a/0.74L) = 0.62$, where $a = 2.36$ Å is the Au(111) layer spacing, $L=6.6$ Å is the electron attenuation length,[43] and the factor of 0.74 applies to the LEED-type energy analyser used.[49]

This gives a simulated signal from just the dissolved O of $(Totd/S)\sum_1^S \exp(-S\,a/0.74L) = \delta\, Totd$, where $\delta$ is 0.26, 0.13 and 0.09 for an $S$ of 10, 20 and 30 layers, respectively. Fig. 8 is a plot of the total simulated signal as a function of the amount of oxygen scavenged, and is consistent with the form of the data of Fig. 6 within the experimental uncertainty for $S = 10$. However, the break in the slope of the curves becomes more extreme for the $S = 20$ or 30 cases, making them less probable. The data for Fig. 6 was taken at 85.1 °C whereas that for Fig. 7 was at 52.2 °C, but this should not affect the basic argument.

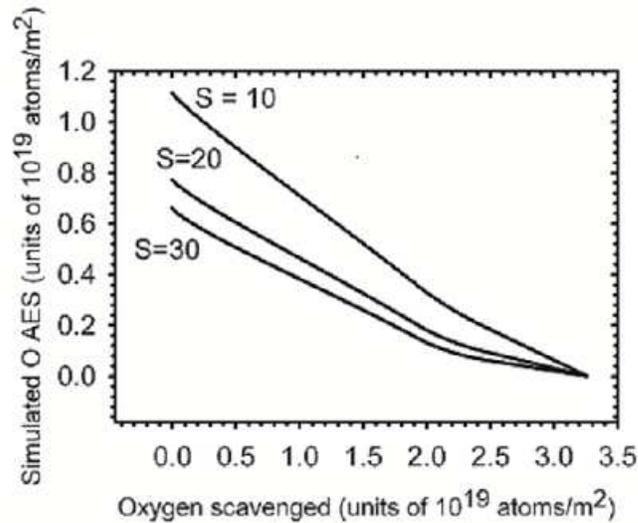

Fig. 8. Simulated form of the oxygen AES amplitude *vs*. the amount of oxygen scavenged, using data from Fig. 7. *S* is the number of Au(111) layers assumed for the selvedge.

## F. Modelling the oxygen uptake curves

The solid black line in Fig. 9 shows another typical uptake curve as a function of time, taken at 85.2 °C and 6 x $10^{-5}$ Torr of $O_2$ as for the curves in Fig. 4. The dotted line $m(t)$ results from an analytical model fitted to the experimental data, as discussed below. The agreement between the data and the model is typical of that obtained for all uptake runs at all temperatures used. $m_o(t)$ and $m_d(t)$ are the fitted areal densities of O atoms in the oxide and dissolved, respectively.

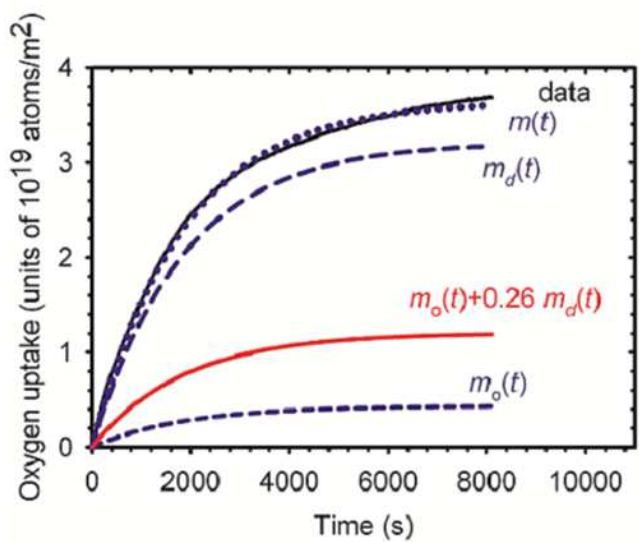

Fig. 9. (Color online) Oxygen uptake and theoretical fit, assuming no CO scavenging, 85.2 °C. Top solid line: experimental. Dots: model fit for total oxygen. Long dashes: dissolved O. Short dashes: O in oxide. Lower solid line: "simulated" AES, $m_o(t)$ + $0.26m_d(t)$.

The similarity in the shapes of the QCM uptake curve and the AES results of Fig. 4 suggest that oxygen adsorbs only at the unoxided region of the surface, either as oxide or as dissolved oxygen atoms. However, the experimental O-uptake curve in Fig. 9 is almost linear at long exposure time. This was typical of all uptake curves if the O exposure time was long enough. As discussed above, this was most likely caused by continuous scavenging of the oxide "cap" by background CO, which allowed more oxygen to enter the Au film as dissolved oxygen. Therefore, the total oxygen content would continually increase almost linearly in time, even when the oxide coverage was nearly complete.

This situation is considerably more complicated to model than the CO scavenging, because two processes, O uptake and CO scavenging, are occurring simultaneously, and the pressure of the atomic oxygen gas and of the background CO are unknown. We know that, at high oxide coverage, the probability of scavenging from the oxide varies as the perimeter of holes in the oxide. No simple model could be found that adequately fitted the oxygen uptake curves, including the linear portion at long exposures, and still gave physically reasonable values for all the fitting parameters. Therefore, a model was used that ignores CO scavenging, which will be relatively weak because of the low CO background pressure. The model assumes that both the oxide and dissolved O that are added during a short time increment are proportional to the fraction of unoxided area because no adsorption occurs in the oxided area. It also assumes that the oxygen indiffusion rate is high enough that the dissolved oxygen near the surface does not impede the uptake of more oxygen. (For the model development, see **Supplementary Material**, section C, at (**insert URL**)). This gives a total oxygen areal density

$$m(t) = M_{max}[1 - \exp(-t/\beta)], \qquad (3)$$

where $\qquad M_{max} = M_o(1 + f_{dis}/f_{ox}) \qquad (4)$

and $\qquad \beta = M_o / F_O f_{ox}. \qquad (5)$

Here, $M_o$ is the oxygen areal density on the oxided surface at full coverage, $f_{dis}$ and $f_{ox}$ is the probability that an O atom hitting unoxided surface creates a dissolved O atom or forms gold oxide, respectively, and $F_O$ is the (unknown) oxygen-atom flux density, proportional to the known $O_2$ flux density. $M_{max}$ would be the maximum quantity of oxygen absorbed for a true saturating exponential. Values for the fitting parameters $\beta$ and $M_{max}$ have been obtained by minimizing the rms difference between the experimental uptake data and Eq. (3), using the Nelder-Mead algorithm. Again, this rms difference was divided by the maximum value for absorbed O density from the experimental uptake curve, to give *yrms* which will be about 0.01 for an average 1% deviation between experiment and model results. The results of the specific fit at 85.2 °C are shown in Fig. 9 as the dots, which had a *yrms* value of 0.015, $M_{max} = 3.65 \times 10^{19}$ atoms m$^{-2}$s$^{-1}$ and $\beta = 1853$s, typical values for these runs. These values are reasonable as can be seen from the figure: $\beta$ is approximately the time constant of the experimental uptake curve and $M_{max}$ is close to the maximum experimental areal density of the absorbed oxygen. The average values for *yrms* and the fitted parameters at each temperature used for the uptake runs, and their standard deviations from the mean, are given in Table II. See **Supplementary Material, section D** (**insert URL**) for the computer codes.

The goodness of this saturating-exponential fit supports the assumption that the oxide forms randomly on the surface, rather than, for example, just at the edges of nucleated oxide islands. The latter process would accelerate as the islands expanded, and then slow after coalescence, and this is not observed in Fig. 9.

Table II. Fitting parameters for oxygen uptake. The numerical factors in the top row give the units for the corresponding column.

| Temp. | No. of runs | | yrms $10^{-2}$ | $M_{max}$ $10^{19}$ m$^{-2}$s$^{-1}$ | $\beta$ s | $f_{dis}/f_{ox}$ | $F_O f_{ox}$ $10^{15}$ m$^{-2}$s$^{-1}$ |
|---|---|---|---|---|---|---|---|
| 52.2 C | 5 | Avg. | 2.16 | 3.28 | 1654 | 6.48 | 2.73 |
|  |  | St. dev. | 0.74 | 0.20 | 325 | 0.45 | 0.56 |
| 85.2 C | 7 | Avg. | 1.44 | 3.46 | 1609 | 6.83 | 2.75 |
|  |  | St. dev. | 0.51 | 0.11 | 210 | 0.25 | 0.37 |
| 101.1 C | 6 | Avg. | 2.25 | 3.41 | 1686 | 6.77 | 2.60 |
|  |  | St. dev. | 0.30 | 0.16 | 172 | 0.36 | 0.29 |
| 120.6 C | 2 | Avg. | 2.38 | 3.49 | 1904 | 6.96 | 2.28 |
|  |  | St. dev. | 0.20 | 0.21 | 23 | 0.47 | 0.03 |

If we again use the value of $M_o = 0.3125 N_A u$,[22] Eqs. (4) and (5) give the values of $f_{dis}/f_{ox}$ and $F_O f_{ox}$ listed in Table II, which should be useful for comparison with values from first-principles calculations. If the partial pressure of the adsorbing gas were known, as would normally be the case, then the last two columns would give the value of both $f_{ox}$ and $f_{dis}$. The average values of the fitted parameters from different sample temperatures agree within the combined standard deviations, showing there is no measurable temperature dependence over this temperature range.

If the indiffusion were slow enough to limit dissolution of more oxygen into the sample, it would be straightforward to include diffusion explicitly as in the 3-parameter fit to the scavenging data. Models were tested that included indiffusion to see if it could explain the linear tail at long exposure times; however, these models still gave flat, exponential-like saturation at long $O_2$ exposure times.

To check that this model also satisfies the requirement of agreeing with the AES data in Fig. 4, the form of the AES data can be simulated using the same approach as discussed for the data from CO scavenging; i.e., a plot of $m_o(t) + 0.26\, m_d(t)$ should have the same time dependence as the experimental AES data in Fig. 4. This is automatically satisfied because the expressions for $m(t)$, $m_o(t)$ and $m_d(t)$ all have the same time dependence (see **Supplementary Material**, section C, at (**insert URL**)), so any linear combination of $m_o(t)$ and $m_d(t)$ can be scaled to agree exactly with $m(t)$, which is clearly close enough to the QCM uptake data to be consistent with the AES data points in Fig. 4. The plot of $m_o(t) + 0.26\, m_d(t)$ is included in Fig. 9. In this simulation, roughly two-thirds of the AES signal comes from the dissolved oxygen.

Note that the linear dependence from CO scavenging at near-saturation cannot be addressed by just adding a term linear in time to Eq. (3), with one more fitting parameter, as this does not take into consideration the CO scavenging proportional to the perimeters of the oxide islands. The perimeters will be changing throughout the oxygen uptake in a very non-linear fashion with time as oxide islands coalesce.

We now return to the question of whether the subsurface oxygen could have formed an oxide, rather than just dissolve in the Au as assumed so far. As stated earlier, an STM and XPS study has shown that at 400 K, oxidation of the Au(111) surface forms large 2D islands consisting of a single oxide.[21] Additional evidence follows. (1) The subsurface oxide would have to be more than one layer thick to accommodate more than 2 MLE of oxygen. One would expect the formation of both a surface oxide and a thickening subsurface oxide to require a more complex kinetic model than the simple "saturating exponential" developed, at least if the subsurface oxide limited O-atom indiffusion. (2) It seems unlikely that the scavenging of surface oxide by the background CO during oxygen uptake would have resulted in the observed linear tail at near-saturation coverage if the extra oxygen was forming a subsurface oxide. Rather, one would expect that the subsurface oxide would thicken to the point at which no more oxygen could penetrate it. (3) A subsurface oxide will form only if it is more stable energetically than dissolved oxygen. Therefore, it seems unlikely that a subsurface oxide would decompose to form dissolved oxygen which could migrate to the surface to be scavenged by CO.

# IV. ESTIMATE OF THE OXIDE THERMAL DESORPTION ENERGY FROM AES MEASUREMENTS

After near-saturation adsorption of oxygen for runs below 120 °C, for which there was little thermal desorption, the sample temperature was raised rapidly to a range of fixed temperatures $T_d$ between 130.5 and 197.8 °C. Oxygen Auger spectra were recorded at regular intervals at each temperature. At 130.5 °C complete desorption took 20 h whereas at 197.8 °C it took 35 min. The thermal desorption at these temperatures also removed dissolved O, as there was no change in the QCM frequency on exposure to CO gas after the vacuum anneal. It was not possible to measure the desorption of oxygen with the QCM in these experiments: the frequency of the quartz crystal was not stable enough with temperature that far from the $f(T)$ turning point, and the time to stabilize the frequency at a substantially higher temperature (at least an hour even at temperatures below 120 °C) occupied too much of the desorption time.

There was no electron-induced desorption of the surface oxygen during AES measurements over 1 hr, so this is not a factor in these experiments. The oxygen Auger amplitudes were measured as described earlier. These amplitudes were plotted as a function of the desorption time after the sample reached $T_d$ and the plots were fitted linearly. The natural logarithms of the slopes were then plotted against $1/T_d$ (K$^{-1}$), as shown in Fig. 10. Assuming the desorption rate varies as $\exp(-E_o/k_B T_d)$, the slope of this plot gives a desorption free energy $E_o$ of 0.9 +/- 0.1 eV where the uncertainty is the standard error in the slope. Applying a correction for the scavenging of oxygen from background CO would decrease the slope of the AES amplitudes as a function of desorption time by about 10% for the lowest temperature, but this is a negligible correction to the logarithm in Fig. 10. The value of $E_o$ from Fig. 10 is significantly lower than the 1.4 eV obtained from TPD from Au(110) in the coverage regions which showed either first- or second-order desorption kinetics for $O_2$.[13] However, TPD will measure both oxide and dissolved oxygen, and our modelling of the oxygen uptake indicates that dissolved oxygen should dominate the TPD spectrum whereas the Redhead analysis[20] used in Ref. 10 applies to surface species only. Moreover, the straight-line fit in Fig. 10 falls well outside the error bars of the data points, which suggests that the process of thermal desorption of oxygen from the oxide may even be changing its mechanism with temperature.

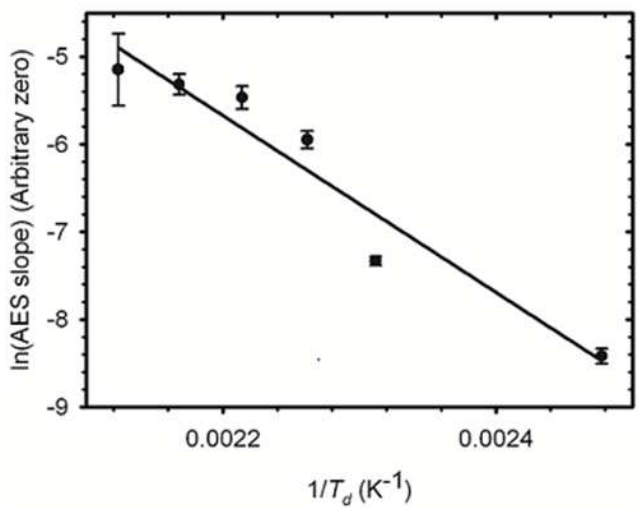

Fig. 10. Arrhenius plot for the oxide thermal desorption energy. The error bars are calculated as $d\ln(y) = dy/y$ where $y$ is the slope of the plot of AES amplitude *vs.* $t$ and $dy$ is the standard deviation of the slope.

As discussed earlier, most researchers found TPD spectra whose peak temperatures were stationary with increasing initial coverage over 1.5 MLE, implying first-order desorption rather than the expected second-order (associative) desorption of $O_2$. A stationary TPD peak might be explained by peaks that are dominated by dissolved oxygen which feeds the surface continuously by out-diffusion over the time of the TPD temperature ramp.

## V. CONCLUSIONS

A high-stability QCM has been used to measure the total absolute mass of absorbed atomic oxygen in gold, both as surface oxide and dissolved oxygen, at an $O_2$ pressure of 6 x $10^{-5}$ Torr and sample temperatures from 50 to 120 °C. The QCM was also used to study the scavenging of the oxygen by CO gas. Stabilizing the sample temperature by controlling the RTD feedthrough temperature, and implementing a correction for changes in room temperature, were crucial for obtaining reliable frequency-time curves. Conditioning the gold by pre-absorption of the gas was necessary to obtain reproducible results for oxygen absorption, and this effect persisted for at least 30 min but not for a day. Presumably this effect was due to a metastable rearrangement of the Au atoms that facilitates in-diffusion. The QCM technique is particularly useful when combined with surface-sensitive techniques such as AES, as complementary bulk and surface information can be obtained in the same apparatus.

Kinetic models, for both oxygen uptake and CO scavenging of oxygen, were developed that were important for understanding the experimental data. Fitting these models to the QCM data also provided kinetic parameters which should be useful for comparison with first-principles calculations. For Au exposed to atomic O, oxide formation and oxygen in-diffusion occur simultaneously with the oxygen uptake limiting as the oxide cover nears completion. An acceptable fit to the data was obtained by assuming the O uptake rate is limited primarily by the amount of oxide already on the surface; the O indiffusion rate is high enough that it is not a limiting factor in the oxygen uptake. The scavenging of oxygen by CO gas has been modelled successfully by assuming that scavenging of the oxide takes place only at the peripheries of holes in the oxide. There was no discernible temperature dependence of either the oxygen uptake, or the scavenging by CO, over the range 50 – 120 °C available with the current quartz crystals. However, this range could be extended by using crystals with $f(t)$ turning points at higher temperatures. The reproducibility of the experimental results and the ability to model them provide a proof-of-principle for the QCM technique for studying the kinetics of the adsorption and diffusion of a gas in a metal.

# ACKNOWLEDGMENTS

Thanks to Bill Atkinson for assistance with computing, to Mark Gallagher for comments on the paper, and to one of the referees for pointing out references 8 and 9. The calculations were performed at the Shared Hierarchical Academic Research Computing Network (SHARCNET: www.sharcnet.ca) and Compute/Calcul Canada. The financial support of NSERC Canada and Trent University is gratefully acknowledged.

**Tables**

Table I. Values of *yrms, W_aox, adis, fi* for CO-scavenging runs after maximum O uptake, at various sample temperatures and CO pressures. The numerical factors in the top row give the units for the corresponding column.

| T °C | Run | $P_{CO}$ $10^{-5}$ Torr | yrms $10^{-3}$ | W_aox $10^{-5}$ m | adis $10^{-3}$ | fi Hz | v $10^{-3}$ $R_c$/s | v/$P_{CO}$ $10^2$ $R_c$/ Torr-s |
|---|---|---|---|---|---|---|---|---|
| 52.2 | Jul28 | 2.5 | 5.86 | 12.3 | 10.7 | 21.2 | 2.72 | 1.09 |
| | | | **5.74** | **12.4** | **10.6** | **infinite** | **2.74** | **1.10** |
| 52.3 | Aug04 | 12.1 | 5.27 | 7.61 | 15.4 | 0.61 | 8.11 | 0.67 |
| | | | **7.69** | **11.0** | **4.11** | **infinite** | **11.7** | **0.97** |
| 53.0 | Aug26 | 3.2 | 5.52 | 9.79 | 8.12 | 2.8 x $10^4$ | 2.76 | 0.86 |
| | | | **5.43** | **9.86** | **8.05** | **infinite** | **2.78** | **0.87** |
| 101.1 | Aug05 | 11.0 | 6.83 | 11.5 | 8.20 | 7.0 | 11.2 | 1.02 |
| | | | **6.73** | **12.1** | **7.22** | **infinite** | **11.7** | **1.06** |
| 101.1 | Aug08 | 10.9 | 8.22 | 11.8 | 8.21 | 2.6 x $10^4$ | 11.3 | 1.04 |
| | | | **7.74** | **12.2** | **7.93** | **infinite** | **11.7** | **1.07** |
| 101.1 | Aug09 | 13.6 | 4.11 | 9.15 | 35.4 | 1.2 | 11.0 | 0.81 |
| | | | **12.4** | **15.1** | **6.97** | **infinite** | **18.0** | **1.32** |
| 101.1 | Aug25 | 3.4 | 5.45 | 13.8 | 13.9 | 1.2 | 4.08 | 1.20 |
| | | | **5.83** | **14.9** | **10.3** | **infinite** | **4.41** | **1.29** |
| | **Average** | | **7.6** | **12.9** | **7.7** | **infinite** | | **1.14** |
| | **St. dev.** | | **2.2** | **1.9** | **1.9** | **infinite** | | **0.17** |

Table II. Fitting parameters for oxygen uptake. The numerical factors in the top row give the units for the corresponding column.

| Temp. | No. of runs | | yrms $10^{-2}$ | $M_{max}$ $10^{19}$ m$^{-2}$s$^{-1}$ | β s | $f_{dis}/f_{ox}$ | $F_O f_{ox}$ $10^{15}$ m$^{-2}$s$^{-1}$ |
|---|---|---|---|---|---|---|---|
| 52.2 C | 5 | Avg. | 2.16 | 3.28 | 1654 | 6.48 | 2.73 |
| | | St. dev. | 0.74 | 0.20 | 325 | 0.45 | 0.56 |
| 85.2 C | 7 | Avg. | 1.44 | 3.46 | 1609 | 6.83 | 2.75 |
| | | St. dev. | 0.51 | 0.11 | 210 | 0.25 | 0.37 |
| 101.1 C | 6 | Avg. | 2.25 | 3.41 | 1686 | 6.77 | 2.60 |
| | | St. dev. | 0.30 | 0.16 | 172 | 0.36 | 0.29 |
| 120.6 C | 2 | Avg. | 2.38 | 3.49 | 1904 | 6.96 | 2.28 |
| | | St. dev. | 0.20 | 0.21 | 23 | 0.47 | 0.03 |

**Figure Captions**

Fig. 1. Oxygen AES spectra. Upper curve: O on Au with the background from clean Au subtracted; lower curve: $Al_2O_3$. The peak minima are at 517 eV for O on Au, and 507 eV for $Al_2O_3$.

Fig. 2. (Color online) Three oxygen uptake curves for a sample held at 85.2 °C. The Apr 27 and Apr 29 data were taken immediately following ion cleaning. The May 31 data were taken 1 day after ion cleaning.

Fig. 3. (Color online) Change in frequency during oxygen uptake at 6 x $10^{-5}$ Torr. The top curve includes a conditioning absorption. The lower curve is without any conditioning absorption.

Fig. 4. (Color on line) Lines: Oxygen uptake at 85.2 °C, measured by the QCM. Dots: Auger amplitude during oxygen uptake. The Jun23 line almost coincides with the Jun22 one. The Jun21 line was for no pre-conditioning.

Fig. 5. (Color online) Complete CO scavenging of oxygen after uptake at 52.2 °C. The CO was admitted at 0 s. The right-hand (blue) curve is obtained from the 121-$\mu$Torr one by scaling the time axis by 121/25 and normalizing the starting amplitude to the 25-$\mu$Torr value.

Fig. 6. AES amplitudes as a function of the amount of oxygen scavenged, at 85.1 °C.

Fig. 7. (Color online) Data and fit for full CO scavenging of oxygen at 52.2 °C and $P$(CO) = 2.5 x $10^{-5}$ Torr. Solid line (black): experimental data. Dots: overall fit. Short dash: O in oxide. Long dash: total dissolved O.

Fig. 8. Simulated form of the oxygen AES amplitude *vs.* the amount of oxygen scavenged, using data from Fig. 7. *S* is the number of Au(111) layers assumed for the selvedge.

Fig. 9. (Color online) Oxygen uptake and theoretical fit, assuming no CO scavenging, 85.2 °C. Top solid line: experimental. Dots: model fit for total oxygen. Long dashes: dissolved O. Short dashes: O in oxide. Lower solid line: "simulated" AES, $m_o(t) + 0.26 m_d(t)$

Fig. 10. Arrhenius plot for the oxide thermal desorption energy. The error bars are calculated as $d\ln(y) = dy/y$ where *y* is the slope of the plot of AES amplitude *vs. t* and d*y* is the standard deviation of the slope.